\documentclass[preprint,12pt]{elsarticle}

\usepackage{amssymb}
\usepackage{amsmath}
\usepackage{xcolor} 
\usepackage[normalem]{ulem} 

\usepackage[utf8]{inputenc}  

\usepackage{lineno}

\journal{Information Fusion}

\begin{document}

\begin{frontmatter}



\title{Explainable Natural Language Processing for Corporate Sustainability Analysis}


\author[1,2]{Keane Ong} 
\author[3]{Rui Mao}
\author[4]{Ranjan Satapathy} 
\author[4]{Ricardo Shirota Filho}
\author[3]{Erik Cambria}
\author[2,5,6]{Johan Sulaeman}
\author[1,2,5,7]{Gianmarco Mengaldo\corref{cor1}}
\cortext[cor1]{corresponding author}

\affiliation[1]{organization={College of Design and Engineering, National University of Singapore},
            addressline={9 Engineering Drive 1}, 
            postcode={117575}, 
            country={Singapore}}

\affiliation[2]{organization={Asian Institute of Digital Finance, National University of Singapore},
            addressline={Innovation 4.0, 3 Research Link, \#04-03}, 
            postcode={117602}, 
            country={Singapore}}

\affiliation[3]{organization={College of Computing and Data Science, Nanyang Technological University},
            addressline={50 Nanyang Ave}, 
            postcode={639798}, 
            country={Singapore}}
            
\affiliation[4]{organization={Institute of High Performance Computing, Agency for Science, Technology
and Research},
            addressline={Fusionopolis Way, \#16-16 Connexis}, 
            postcode={138632}, 
            country={Singapore}}

\affiliation[5]{organization={Sustainable and Green Finance Institute, National University of Singapore},
            addressline={Innovation 4.0, 3 Research Link, \#02-02}, 
            postcode={117602}, 
            country={Singapore}}

\affiliation[6]{organization={NUS Business School, National University of Singapore},
            addressline={Mochtar Riady Building, 15 Kent Ridge Dr, Mocthar Riady Building}, 
            postcode={119245}, 
            country={Singapore}}

\affiliation[7]{organization={Honorary Research Fellow, Imperial College London},
}

\begin{abstract}
Sustainability commonly refers to entities, such as individuals, companies, and institutions, having a non-detrimental (or even positive) impact on the environment, society, and the economy. With sustainability becoming a synonym of acceptable and legitimate behaviour, it is being increasingly demanded and regulated.
Several frameworks and standards have been proposed to measure the sustainability impact of corporations, including United Nations' sustainable development goals and the recently introduced global sustainability reporting framework, amongst others. However, the concept of corporate sustainability is complex due to the diverse and intricate nature of firm operations (i.e. geography, size, business activities, interlinks with other stakeholders). As a result, corporate sustainability assessments are plagued by subjectivity both within data that reflect corporate sustainability efforts (i.e. corporate sustainability disclosures) and the analysts evaluating them. This subjectivity can be distilled into distinct challenges, such as incompleteness, ambiguity, unreliability and sophistication on the data dimension, as well as limited resources and potential bias on the analyst dimension. Put together, subjectivity hinders effective cost attribution to entities non-compliant with prevailing sustainability expectations, potentially rendering sustainability efforts and its associated regulations futile. To this end, we argue that Explainable Natural Language Processing (XNLP) can significantly enhance corporate sustainability analysis. Specifically, linguistic understanding algorithms (lexical, semantic, syntactic), integrated with XAI capabilities (interpretability, explainability, faithfulness), can bridge gaps in analyst resources and mitigate subjectivity problems within data.
\end{abstract}





\end{frontmatter}

\section{Introduction}

Sustainability, intended as having a non-detrimental impact on the environment, society, and the economy, is becoming increasingly essential for humanity's future. 
Various efforts are being undertaken at different levels of the societal and economic hierarchy, from country-level to institutional- and business-level entities, and in some cases, down to individuals. 
These efforts are focused on making these entities \textit{sustainable}, amid reputational risks~\cite{NIKOLAOU2015499}, and pressing challenges including global climate change, widespread inequity, governance malpractices, and geopolitical instability~\cite{folque2021sustainable}.   

In this work, we focus on sustainability analyses of entities that provide sustainability reports (also referred to as sustainability disclosures); in particular public companies and institutions. These entities are crucial components in the sustainable development~\cite{corporatesusimpt}, and we refer to their sustainability as \textit{corporate sustainability}. 


Given that these entities have a substantial stake in global sustainability, the analysis of their sustainability, or corporate sustainability analysis, is critical~\cite{rajesh2020exploring}. Yet, this is extremely \textit{challenging} due to the inherent complexity of corporate sustainability as a concept. 
To elaborate, the complexities within firms (e.g., geography, size, business activities) and outside the firms (e.g., their global supply chains), as well as the firms' relationships with non-business stakeholders, lead to the evolving frameworks and guidelines put forward to address corporate sustainability issues~\cite{buallay2020sustainability,derqui2020towards}. Further complicating the matter is the lack of globally mandated standards for sustainability reporting~\cite{esgdivergence}.

For example, a firm may use a convenient corporate sustainability reporting framework that magnifies only certain aspects of a firm's practices~\cite{Darnall2022}, while omitting key sustainability dimensions~\cite{demastus2024organizational}. 
Or, within a given regulatory disclosure framework that a firm shall abide to, it may provide sustainability disclosures that are difficult for external parties to digest, and whose integrity and transparency may be questionable~\cite{Boiral2015}. 


This first aspect, that we label the \textit{data dimension}, is inherently \textit{subjective}, as it depends on what the firm chooses to disclose and how. However, this data dimension is only one side of the challenge. 

Indeed, today's corporate sustainability analysis is carried out by human analysts who read sustainability disclosures provided by companies (i.e., the data dimension), and provide corporate sustainability evaluations based on a combination of sustainability frameworks (e.g., Environmental, Social, Governance (ESG)~\cite{Dmuchowski2023}, Global Reporting Initiative (GRI)~\cite{Machado2021}, Sustainability Accounting Standards Board (SASB)~\cite{Eng2022}, Greenhouse Gas Protocol and Carbon Disclosure Project (CDP)~\cite{Pitrakkos2020}, United Nations' Sustainable Development Goals (SDG)~\cite{pradhan2017systematic}, United Nation Global Compact (UNGC)~\cite{orzes2018united}). These evaluations require an extensive amount of human hours~\cite{Ni2023}, and yet are inevitably influenced by the inherent biases in the frameworks adopted as well as the analysts' own subjectivity~\cite{buyukozkan2018sustainability, anlaystbias}. 

This second aspect, that we label the \textit{analyst dimension}, is also inherently \textit{subjective}, similar to the \textit{data dimension}. 

Amid the complexities of corporate sustainability, analysts face the challenge of dealing with \textit{subjective} (and likely flawed) data, and they need to account for their own \textit{subjective} perspective. The interplay between the \textit{data} and \textit{analyst} dimensions is critical, as it results in a highly subjective evaluation of corporate sustainability that would hamper efforts to achieve corporate sustainability goals. 



We argue that, to address the subjectivity inherent to data and analysts, it is crucial to adopt artificial intelligence (AI), specifically through linguistic understanding methods enhanced by explainable AI (XAI) capabilities, or what is termed as explainable natural language processing (XNLP)~\cite{xnlp}. To qualify, adopting XNLP entails \textit{complementing} human analysts instead of replacing them. Within corporate sustainability analysis, analysts should remain `in the loop' -- while XNLP can expeditiously process numerous sustainability disclosures to unveil consistent and meaningful insights, analysts can further synergise these insights to inform their corporate sustainability assessments.   

\begin{figure*}[hbt!]
    \centering\includegraphics[width=\textwidth]{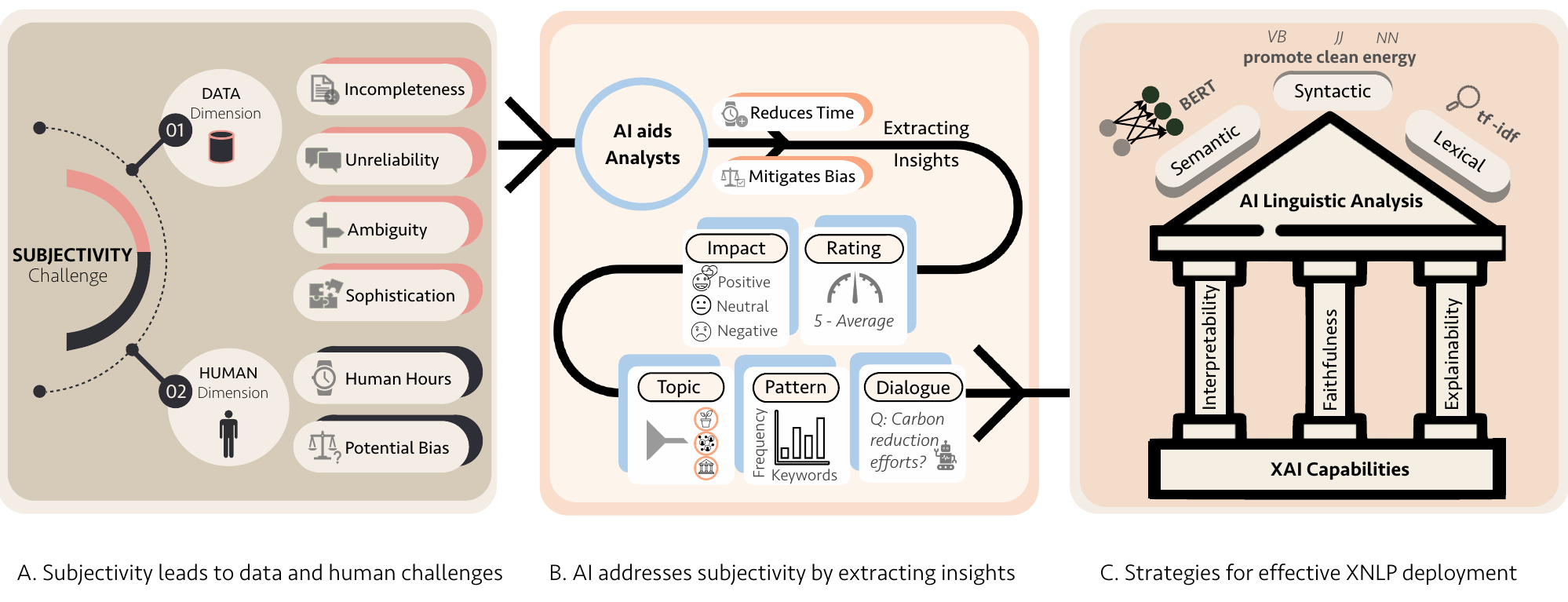}
    \caption{Proposed framework for XNLP enhanced sustainability analysis}
    \label{fig:Overview}
\end{figure*}

In this paper, we follow the three foundational building blocks presented in Fig.~\ref{fig:Overview}, focusing on how XNLP can pave the way for effective corporate sustainability analysis by addressing the \textit{subjectivity} issues that plague the field. We first break down the subjectivity challenge into its \textit{data} and \textit{analysts} dimensions (Fig.~\ref{fig:Overview}(A)), characterising the problems for XNLP to address. Next, we consolidate NLP's distinct advantages for solving these challenges by delineating its specialised applications (Fig.~\ref{fig:Overview}(B)). Finally, we lay the groundwork for effective XNLP deployment within this domain (Fig.~\ref{fig:Overview}(C)). We realise this by surveying NLP methods to match specific applications based on their different levels of linguistic understanding, and proposing strategies to integrate XAI capabilities. By offering insights that blend technical depth with broader viewpoints of sustainability, we pave the way for XNLP's adoption within corporate sustainability analysis. 

\section{The subjectivity challenge}

As a result of the \textit{complexity} inherent to corporate sustainability, \textit{subjectivity} commonly underpins corporate sustainability analysis. Specific challenges arising from subjectivity can be distilled into the \textit{data} and \textit{analyst} dimensions. We expound on these dimensions to frame (X)NLP's suitability for corporate sustainability analysis.

\subsection{The data dimension}

Data is the first building block, or entry point, of corporate sustainability analysis, as depicted in (Fig.~\ref{fig:Overview}(A)). While the data required for corporate sustainability assessments can come from various sources, \textit{corporate sustainability disclosures} are a predominant source~\cite{nguyen2020empirical}. These disclosures consist of reports released by businesses detailing their own sustainability efforts~\cite{nguyen2020empirical}. For example, \textit{general sustainability reports} typically follow established sustainability frameworks to describe a company's sustainability efforts and achievements~\cite{abeysekera2022framework}, and are typically accompanied by \textit{integrated reports} that aggregate the general sustainability information with financial implications~\cite{fasan2013annual}. 


Additional data may come from \textit{government-mandated sustainability disclosures}, where companies are required to release specific aspects of their sustainability performance~\cite{tan2022assembling}, such as their environmental impact~\cite{Scott2022}, or ESG (Environmental, Social, and Governance) practices~\cite{SECsusdiclosures}. Data originating from NGO sources and media coverage may also be relevant to assess a company's sustainability efforts~\cite{WONG2022101045, ngoimpactinfo}. Yet these data are influenced by sustainability disclosures~\cite{materialityanalysismedia2020, fiaschi2020bad}, underlining the importance of focusing our discussion on the sustainability disclosures themselves. 

Due to the intricacies and multi-faceted dimensions of corporate sustainability, these data sources inherently contain \textit{subjectivity}. They involve competing notions of corporate sustainability, expressed through a vast and diverse array of sustainability frameworks~\cite{complexitydiffframeworkruiz, Meuer2020}. Moreover, even within a specific framework, deciding what information to convey necessitates a value judgement (e.g., financial vs. environmental materiality), further exacerbating the partiality involved~\cite{beske2020materiality}. We argue that this \textit{subjectivity} at the data level manifests itself in four data grand challenges that analysts need to face, namely \textit{incompleteness}, \textit{unreliability}, \textit{ambiguity}, and \textit{sophistication}. These 4 data challenges are reported on the first block of (Fig.~\ref{fig:Overview}(A)), referred to as \textit{data dimension}. We detail them in the following.

\begin{description}
\item \textbf{\textit{Incompleteness}}. It refers to how data may not provide a comprehensive and complete view of a firm's sustainability efforts. As certain sustainability disclosures are voluntary, companies can decide on what is material enough to release as sustainability information, thereby raising concerns about openness and the omission of important data~\cite{beske2020materiality}. 
%
\item \textbf{\textit{Unreliability}}. It refers to the trustworthiness and accuracy of sustainability data. Indeed, corporate and government-linked sustainability disclosures may involve greenwashing, where organisations misleadingly claim and exaggerate that their activities or products are more environmentally friendly than they really are~\cite{de2020concepts}. Or they use non-transparent carbon offset products, that often do not deliver the offsets promised~\cite{delacote2024strong}. In fact, such developments are particularly frequent for firms that are of larger sizes, to contend with increased dealing with stakeholders~\cite{greenwasharticle}. 
\item \textbf{\textit{Ambiguity}}. It refers to data being vague and unclear, making negative content less conspicuous~\cite{RePEc:inm:ororsc:v:30:y:2019:i:6:p:1207-1231}. One example is sharing information without accompanying it with the appropriate context~\cite{higgins2020managing}. This may lead analysts to misunderstand the sustainability efforts of companies. Ambiguity is related, to some extent, to greenwashing~\cite{zharfpeykan2021representative}, although it does not explicitly refer to inaccurate or misleading data as unreliability does.  
\item \textbf{\textit{Sophistication}}. It refers to sustainability disclosures being tediously lengthy, text-heavy, and requiring specialised knowledge to comprehend, making them an extremely sophisticated writing category~\cite{smeuninx2020measuring}. As a result, firm sustainability reports can require many hours of human effort to understand, let alone derive useful insights~\cite{goel2020mining}. 
\end{description}
%

\subsection{The analyst dimension}
Human analysts must synthesise and interpret sustainability data to produce accurate corporate sustainability assessments. To describe this process, relevant data is collected according to the objectives and scope of the corporate sustainability analysis (i.e. evaluating all operations versus one area). Thereafter, the analysis entails leveraging sustainability frameworks or metrics (i.e., carbon footprint) to measure different impact dimensions (environmental, social, economic etc.)~\cite{tsalis2020new}. The costs and inaccuracies of this process are exacerbated by human limitations. Specifically, human analysts cannot deal well with the challenges of \textit{subjective} data, and are not free from \textit{subjective} interpretations. These human-centric issues are encapsulated by \textit{limited human-hours} and \textit{potential bias}.

\begin{description}
\item[\textbf{Limited human-hours}.] Analysts have a limited amount of time to produce corporate sustainability assessments. However, the process entails extracting meaningful information from corporate sustainability disclosures~\cite{goel2020mining}, which requires a significant amount of time i.e., human-hours~\cite{Ni2023}. In particular, the disclosures are extremely time consuming to read and understand due to the \textit{sophistication} and \textit{ambiguity} of the data within them. Moreover, given the potential \textit{incompleteness} of disclosure information, analysts may analyse additional sources to ascertain a firm's sustainability, further increasing the human-hours required~\cite{qorri2022practical}.
    
\item[\textbf{Potential Bias}.] Analysts' interpretations of sustainability data are subject to bias. This is not least due to the complex and wide ranging factors to be considered, such as firm characteristics, evolving sustainability guidelines, and economic implications, amongst others, often leading to leading to varying interpretations amongst analysts~\cite{buyukozkan2018sustainability, anlaystbias}. To further compound the potential for bias, data \textit{ambiguity} and \textit{unreliability} hinders clear interpretation of sustainability information by obfuscating negative content~\cite{RePEc:inm:ororsc:v:30:y:2019:i:6:p:1207-1231}.
\end{description}

\subsection{NLP to the rescue}

\textit{Subjectivity} pervades through the data and analyst dimensions, making corporate sustainability analyses an onerous task. To alleviate this burden, we argue that AI technologies, through linguistic understanding enabled by natural language processing (NLP), can significantly aid analysts. NLP can automatically process hundreds of sustainability disclosures~\cite{stammbach2023environmental}, summarising key details and extracting insights for analysts. This reduces the \textit{human hours} required to analyse disclosures and partially addresses \textit{potential bias} by providing more consistent and replicable insights~\cite{abram2020methods, Chowdhary2020NLP}. We concretise NLP's usefulness in this domain by detailing five NLP tasks for corporate sustainability analysis \textbf{(T1-T5)} that we deem critical, and that have been partially explored in the literature.

\begin{description}
\item \textbf{(T1) Topic extraction}.
Topic extraction allows an analyst to obtain, from sustainability disclosures, content most pertinent to corporate sustainability. Consequently, analysts do not need to sieve through all information, and can focus their analysis on material information~\cite{BINGLER2022102776}. \textit{Thematic} extraction involves classifying sentences within sustainability frameworks. 
This includes sorting text into GRI topics~\cite{polignano-etal-2022-nlp}, ESG pillars~\cite{LEE2023119726}, or ESG-related concepts~\cite{kang-el-maarouf-2022-finsim4}. 
\textit{Universal} extraction filters textual data into generic categories that are separate from sustainability frameworks. 
For instance, labelling environmental claims~\cite{stammbach2023environmental}, emissions targets~\cite{ClimateBERT-NetZero}, or climate relevant information~\cite{BINGLER2022102776}. On the other hand, \textit{topic discovery} determines topics from sustainability disclosures a posteriori. For example, sustainability topics were uncovered from the sustainability disclosures of shipping companies~\cite{Zhou2021}.

\item \textbf{(T2) Impact Classification}. Impact insights can also be derived from sustainability disclosures, providing analysts a valuable perspective on corporate sustainability. 
\textit{Polarity} classification determines the impact polarity of text from sustainability disclosures. 
For instance, within SEC 10k reports, sustainability related sentences can be labelled as positive, negative or neutral in relation to their impact toward sustainability~\cite{finetuneTransformerESG}.
Alternatively, \textit{strategic} classification uncovers a text's prospective impact. Examples include the risk and opportunity impact analysis of ESG-related texts~\cite{tseng2023dynamicesg}.

\item \textbf{(T3) Rating}. A firm's sustainability performance can be automatically scored to expedite an analyst's evaluation process. \textit{Aggregated} rating methods include computing the frequency of sentiment labels for ESG-related headlines~\cite{fischbach2022automatic}, and averaging the E, S, and G topic classification probabilities of documents~\cite{sokolov2021building}. \textit{Direct} methods explicitly score companies without amalgamating already labelled data. For instance, the prediction of ESG risk ratings from sustainability topics~\cite{RiskfactorESGpred}, as well as regressing lexical features to derive ESG risk ratings~\cite{ignatov2023esgregression}. Outside of holistic sustainability frameworks (i.e. ESG), \textit{direct} methods also include estimating carbon emissions from bank transactions~\cite{gonzalez2022explainable}, and constructing sustainability indices from the lexical features of disclosures~\cite{tian2023lexiconriskratingdataset}.


\item \textbf{(T4) Dialogue.} Processing sustainability disclosures into interactive dialogue format allows an analyst a user-friendly way for querying information. For instance, a chatbot integrated with Retrieval-augmented Generation (RAG) can be queried to summarise or answer analysts' questions about disclosures, allowing analysts to extract information without reading the copious amounts of text typically contained within them~\cite{Ni2023,garigliotti2024sdgrag}. By emphasising critical information, dialogue systems mitigate the obfuscation of key information within disclosures, potentially enhancing the accuracy of sustainability analysis~\cite{sinnewe2021obfuscate}.

\item \textbf{(T5) Linguistic Patterns}. Linguistic analysis deepens an analyst's understanding of how a disclosure's style and structure relates with corporate sustainability performance.
For example, keyword frequency and word relationships within oil and gas compliance reports can uncover environmental violation patterns~\cite{Bi2023}, and the syntactic complexity of CSR reports for good CSR performers can also be distinguished~\cite{Clarkson2020}. 

\end{description}
%



\section{NLP methods for linguistic understanding}

Tasks T1--T5 help human analysts decode sustainability disclosures, addressing the domain's \textit{subjectivity} challenge. To successfully accomplish T1--T5, it is useful to comprehend NLP methods according to their different levels of \textit{linguistic understanding} (lexical, semantic, and syntactic). These three categories unpack the complexity of NLP algorithms, from simple lexical evaluation to complex semantic and syntactic processing, providing insight into the suitability of certain methods for specific tasks.
To this end, Tab.~\ref{tab: XAI_CE} presents a structured overview, detailing which tasks, T1--T5, have been addressed in the literature by the three categories of NLP methods introduced.
We delve more into these three different approaches, providing a concise review of what has been already done in the corporate sustainability analysis space.
\begin{table*}[ht!]
\centering
\caption{Classification of NLP papers for corporate sustainability analysis, detailing the algorithm category and tasks covered.}
\label{tab: XAI_CE}
\ 
\resizebox{\textwidth}{!}{
\begin{tabular}{|p{3.5em}|p{0.5em}|p{0.5em}|p{0.5em}|p{1.5em}|p{1.5em}|p{1.5em}|p{1.5em}|p{1.5em}|p{1.5em}|p{1.5em}|p{1.5em}|p{1em}|p{1em}}
\hline
Paper & \multicolumn{1}{c|}{Lexical} & \multicolumn{1}{c|}{Semantic} & \multicolumn{1}{c|}{Syntactic} & \multicolumn{3}{c|}{Topic Extraction} & \multicolumn{2}{c|}{\begin{tabular}[c]{@{}c@{}}Impact \\ Classification\end{tabular}} & \multicolumn{2}{c|}{\begin{tabular}[c]{@{}c@{}}Rating\end{tabular}} & \multicolumn{1}{c|}{Dialogue} & \multicolumn{1}{c|}{\begin{tabular}[c]{@{}c@{}}Linguistic\\ Patterns\end{tabular}} \\ \hline
 &  &  &  & \rotatebox[origin=c]{90}{Thematic} & \rotatebox[origin=c]{90}{Universal} & \rotatebox[origin=c]{90}{Topic Discovery} & \rotatebox[origin=c]{90}{Polarity} & \rotatebox[origin=c]{90}{Strategic} & \rotatebox[origin=c]{90}{Aggregated} & \rotatebox[origin=c]{90}{Direct} & & \\ \hline
~\cite{polignano-etal-2022-nlp} & \checkmark & \checkmark & & \checkmark & & & \checkmark & & & & & \\ \hline
~\cite{LEE2023119726} & & \checkmark & & \checkmark & & & & & & & & \\ \hline
~\cite{stammbach2023environmental} & \checkmark & \checkmark & & & \checkmark & & & & & & & \\ \hline
~\cite{BINGLER2022102776, ClimateBERT-NetZero} & & \checkmark & & & \checkmark & & & & & & & \\ \hline
~\cite{Zhou2021} & \checkmark & & &  & & \checkmark & & & & & & \\ \hline
~\cite{kang-el-maarouf-2022-finsim4} & \checkmark & \checkmark & \checkmark & \checkmark & \checkmark & & & & & & & \\ \hline
~\cite{finetuneTransformerESG}  & \checkmark & \checkmark & & & & & \checkmark & & & & & \\ \hline
~\cite{tseng2023dynamicesg}  & & \checkmark & & \checkmark & & & & \checkmark & & & & \\ \hline
~\cite{fischbach2022automatic}  & \checkmark & \checkmark & & \checkmark & \checkmark & & \checkmark & & \checkmark & & & \\ \hline
~\cite{sokolov2021building}  & & \checkmark & & \checkmark & & & & \checkmark & \checkmark & & & \\ \hline
~\cite{RiskfactorESGpred}  & \checkmark & \checkmark & & & & \checkmark & & & & \checkmark & & \\ \hline
~\cite{gonzalez2022explainable}  & & \checkmark & \checkmark & & & & & & & \checkmark & & \\ \hline
~\cite{mehra2022esgbert}  & & \checkmark & & & & & & & & \checkmark & & \\ \hline
~\cite{goel2020mining} & \checkmark & \checkmark & & \checkmark & & & & & & & & \\ \hline ~\cite{Bi2023} & \checkmark & & & & & & & & & & & \checkmark \\ \hline ~\cite{Clarkson2020} & \checkmark & & \checkmark & & & & & & & & & \checkmark \\ \hline
~\cite{Ni2023,garigliotti2024sdgrag} & & \checkmark & & & & & & & & & \checkmark & \\ \hline~\cite{ignatov2023esgregression, tian2023lexiconriskratingdataset} & \checkmark & & & & & & & & & \checkmark & & \\ \hline

\end{tabular}
}
\end{table*}

\subsection{Lexical methods}

Lexical-based methods focus on the statistical occurrence of keywords and terms within sustainability data, thereby being  relatively simple and easy to interpret. 

More specifically, lexical methods identify lexical terms appearing in text through string matching, and compute their associated frequency (or rate) of occurrence. A formalisation of this framework is given by the term frequency-inverse document frequency (TF-IDF) equations  
\begin{subequations}
\begin{align}
\operatorname{TF-IDF}(t, d, D) &=\operatorname{TF}(t, d) \cdot \operatorname{IDF}(t, D),\label{eq:tf1}\\[0.6em]
\operatorname{TF}(t, d) &=\frac{f_{t, d}}{\sum_{l^{\prime} \in d} f_{l^{\prime}, d}},\label{eq:tf2}\\[0.3em]
\operatorname{IDF}(t, D) &=\log \frac{N}{\{d \in D: t \in d\}},\label{eq:tf3}
\end{align}
\end{subequations}
where TF is the term frequency that quantifies the rate of occurrence of a lexical term $t$ in a document $d$ (e.g., sustainability disclosures), computed as a proportion of all terms in the document -- i.e., Eq.~\eqref{eq:tf2}, while IDF is the inverse document frequency that measures the importance of a term across multiple documents Eq.~\eqref{eq:tf3}. 
A lexical term is less important if it appears in more documents, and vice-versa. 
This prevents words which are commonly used, but have relatively insignificant meaning such as `he, she, they', to be flagged as significant. 
The quantity TF-IDF in Eq.\eqref{eq:tf1} provides a measure of the overall statistical significance of a keyword, by weighting Eq.~\eqref{eq:tf2} and Eq.~\eqref{eq:tf3}. 
The three quantities, TF, IDF, and TF-IDF may be utilised independently from one another, subject to the use case. 

Given their simplicity, lexical methods are appropriate for relatively uncomplicated tasks that do not require significant linguistic understanding. For example, lexical methods can be deployed for task (T5) \textit{linguistic patterns}, to derive keyword (term) frequency patterns in sustainability disclosures~\cite{Bi2023}. On the other hand, for more complex tasks such as T1--T4, lexical analysis is utilised as a feature engineering tool rather than a standalone method. For example, TF-IDF vectorisation has been employed to identify companies of interest for ESG classification from news headlines~\cite{fischbach2022automatic}, while TF-IDF-extracted features have been used to detect environmental claims~\cite{stammbach2023environmental}.

\subsection{Semantic methods}
Semantic methods interpret textual content to understand its meaning~\cite{mao2024semantic}. This can extend beyond understanding the presence of individual words, to involve contextual understanding and implicit connotations. Unlike lexical methods that simply identify word occurrences, semantic approaches often leverage word embeddings, which are vector representations of words within a multi-dimensional space. We further distinguish these two approaches by analysing the sentence ``\textit{Renewable energy is sustainable, as it does not utilise finite resources}". A lexical approach registers the occurrences of individual words (i.e. `renewable', `sustainability', `finite'), without grasping their deeper and interconnected meanings. On the other hand, semantic methods, by determining the proximity of their respective word vectors, interpret `renewable' as synonymous with `sustainability' but antonymous with `finite'. This helps semantic algorithms comprehend that `renewable' is conceptually linked to `sustainability', unlike `finite' which implies that resources are exhaustible.

Traditional methods such as GloVe derive text embeddings from word co-occurrences, and can be leveraged to capture semantic relationships~\cite{pennington2014glove}. For instance, similarity methods can be employed on word embeddings to classify sentences according to their relevance to each ESG indicator~\cite{goel2020mining}. However, while useful, these traditional methods may not capture linguistic subtleties comprehensively, paving the way for more advanced techniques. 

As a significant leap in semantic understanding, deep learning models such as BERT~\cite{devlin2018bert} powerfully capture the intricacies of text. BERT employs bidirectional attention mechanisms to compute the two-way influence between words. As a result, important contextual and semantic information of a sentence can be captured. BERT develops general language understanding through pre-training on a wide variety of text corpora, and subsequently is fine tuned for specific tasks. 

As they attain a high level of semantic understanding, BERT models are suitable for tasks within the corporate sustainability analysis field that require complex language understanding, without being specifically structured for generative tasks. Therefore, BERT is appropriate for complex language understanding tasks like (T1) \textit{topic extraction}, (T2) \textit{impact classification} and (T3) \textit{rating}, but not generative tasks such as (T4) \textit{dialogue}. In line with this, BERT-like models have been used for ESG label classification~\cite{LEE2023119726}, producing company environmental scores~\cite{mehra2022esgbert}, and to derive topics and their associated keywords for ESG-related risk factors~\cite{RiskfactorESGpred} using BERT-generated text embeddings.

Similar to BERT, generative large language models (Gen-LLMs) such as GPT-4 also involve the comprehension of text by exploiting deep learning based attention architectures~\cite{achiam2023gpt}, and pre-training on large text corpora. However,  by learning to predict the next word given the preceding text~\cite{dong2019unified}, their focus extends beyond understanding language, to the generation of coherent text, as well as inference and reasoning capabilities~\cite{inferencellm}. This makes them well-suited for generative answering and summarisation tasks found within (T4) \textit{dialogue}, and advantageous for reasoning tasks like (T3) \textit{rating} which involves complex evaluation~\cite{mao2024gpteval}. As such, generative LLMs have been leveraged for transforming TCFD disclosures into dialogue format, as well as evaluating sustainability disclosures for conformity to TCFD reporting guidelines~\cite{Ni2023}. The latter demonstrates the reasoning abilities of generative LLMs, showing how they can be leveraged to evaluate corporate sustainability performance from disclosures.


\subsection{Syntactic methods}
Syntactic approaches involve analysing the structure of sentences in terms of grammatical and language rules~\cite{zhang2023syntactic}. This contrasts with other approaches like analysing the occurrence of lexicons and interpreting textual meaning. To elaborate on the latter, while semantic methods can involve a higher level of abstraction (i.e., comprehending the hidden or implied connotations of words), syntactic measures focus on the organisation of words and phrases within a sentence without interpreting them more broadly. 

Syntactic methods include part-of-speech (POS) tagging~\cite{chiche2022pos}, dependency parsing~\cite{li2020dependency}, and constituency parsing~\cite{yang2020constituency}, amongst others. These methods can augment efforts to accomplish several complex NLP tasks for corporate sustainability analysis, although they have only seen limited application so far. For instance, when combined with other methods like semantic analysis, text can be analysed more granularly. In other domains such as finance and economics, aspect-based sentiment analysis (ABSA) has already been achieved through POS tagging and semantic rules~\cite{mao2021bridging,consoli2022fine}. Such integrated approaches can more intricately tackle task (T2) \textit{impact classification}, by revealing impact insights toward aspects. To illustrate with an example, within the sentence \textit{``Company X has good emissions performance despite having poor employee wellness"}, ABSA allows for the aspect `emissions performance' to be labelled as positive and the aspect `employee wellness' to be labelled as negative. Distinguishing between the impact of the two aspects enables a finer analysis of sustainability text. Additionally, syntactic analysis can also be used for identifying companies and their relationships with respect to sustainability activities~\cite{automatedESG_jointentity_relation}.   Such methodologies can be enabled by explicit syntactic attributes (i.e., POS, dependency parsing) in addition to semantic patterns. Topics can be derived from the extracted links between companies and their sustainability practices to solve task (T1) \textit{topic extraction}. Additionally, these links can be further processed to evaluate corporate sustainability performance, in the spirit of task (T3) \textit{rating}. Other applications of syntactic analysis include evaluating the grammatical clauses of disclosure sentences~\cite{Clarkson2020}, in line with (T5) \textit{linguistic patterns}, or the parsing of sustainability-related concepts, in a similar vein to the extraction of financial concepts~\cite{du2023finsenticnet}.

\begin{figure*}
    \centering
    \includegraphics[width=0.8\textwidth]{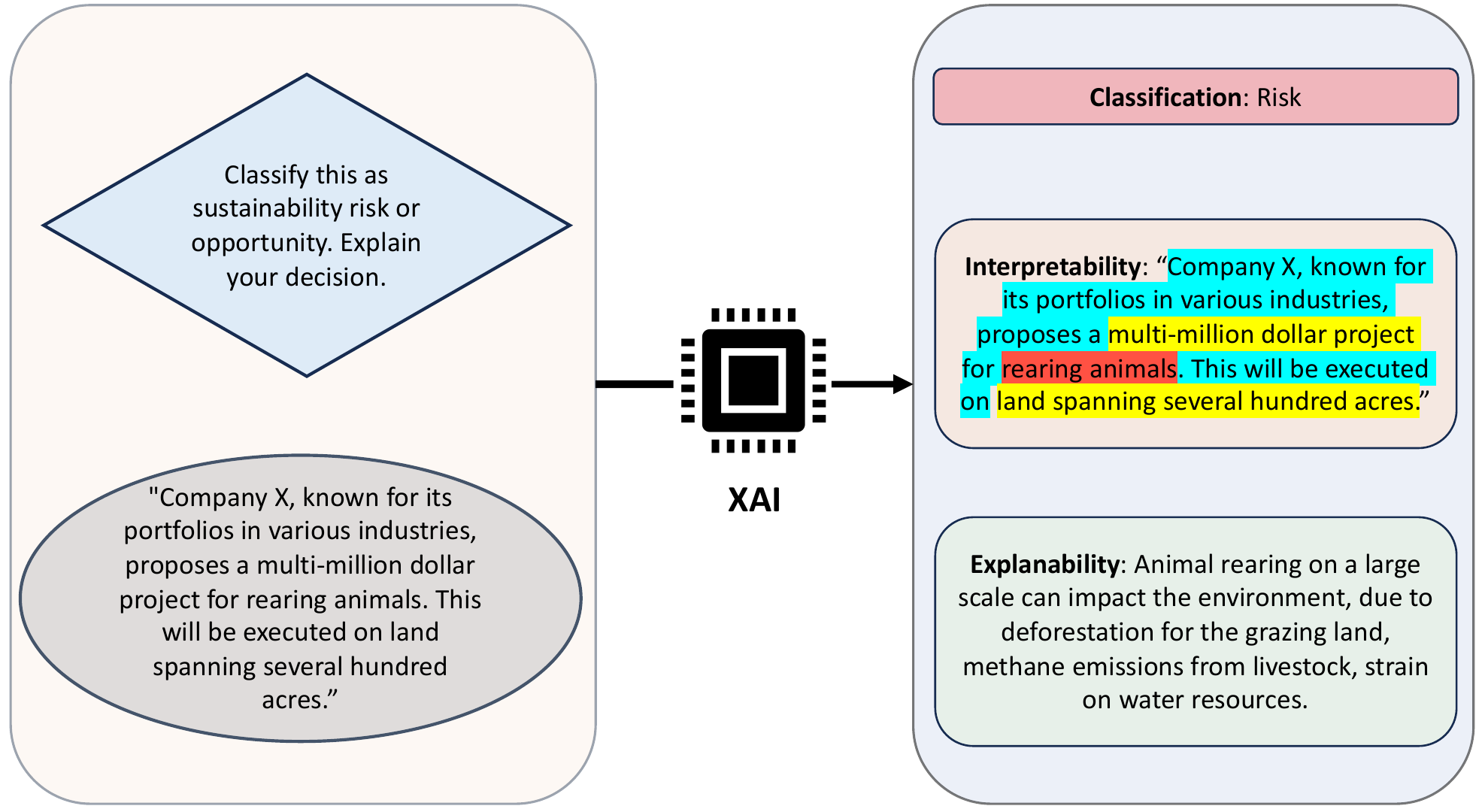}
    \caption{Alongside outputting a decision for sustainability impact classification, XAI can provide an interpretation and explanation}
    \label{fig:interpretability_vs_explain_vs_faithful}
\end{figure*}

\section{XNLP enhances subjectivity mitigation}
The exploration of NLP methods from the \textit{lexical, semantic, syntactic} perspectives provides a pathway for achieving tasks T1-T5, mitigating the \textit{subjectivity} challenges within sustainability disclosures. As we move forward, our focus shifts toward enhancing these efforts by scaffolding the NLP methods with XAI capabilities (\textit{interpretability, explainability and faithfulness}), forming the core components of explainable natural language processing (XNLP) as defined earlier. Despite their potential, these XAI features are still uncommon within the domain of NLP for corporate sustainability analysis, representing a promising research area that has yet to gain traction. We describe these capabilities below. 

\begin{description}
\item \textbf{\textit{Interpretability}}. An AI model's capacity to provide an understanding of its mechanism~\cite{sevenpillars}. For example, scoring which keywords are most salient for an algorithm's classification~\cite{schuff2022human}.

\item \textbf{\textit{Explainability}}. An AI model's capacity to explain why it produces an output~\cite{sevenpillars}. For example, explaining the reasoning steps for model decisions, as enabled by large language models~\cite{jie2024interpretable}. 

\item \textbf{\textit{Faithfulness}}. The accuracy of an AI model's \textit{interpretability} and \textit{explainability} with respect to its true workings~\cite{liu2022rethinking,turbe2023evaluation,wei2024revisit}. For example, the extent to which a model's feature salience scores and provided explanations are representative of its actual workings. Analysing \textit{faithfulness} can be challenging given the lack of ground truth for a model's explainability or interpretability, with several approaches involving dataset modification to verify the feature importance estimated by interpretability methods~\cite{hooker2019roar}.

\end{description} 


Integrating XAI capabilities extends the effectiveness of NLP methods, ensuring they further address the \textit{subjectivity} challenges of the data and analyst dimensions. This ultimately enhances NLP-driven corporate sustainability analysis. We develop this idea by elaborating on the value of these XAI capabilities for NLP-driven corporate sustainability analysis.

\subsection{Achieving trustworthy insights}
XAI capabilities allow NLP models to extend beyond classifications and predictions, providing analysts useful insights through faithful explanations and interpretations~\cite{lipton2018mythos,mengaldo2024explain}. These insights enhance NLP trustworthiness within corporate sustainability analysis, by clarifying  model decisions along with the \textit{ambiguity} of sustainability disclosures. 


For example, figure~\ref{fig:interpretability_vs_explain_vs_faithful} highlights how \textit{interpretability, explainability} and \textit{faithfulness} enhances trust in an NLP model's classification for sustainability text. \textit{Interpretability} allows an analyst to verify that appropriate and relevant text features (i.e. `rearing animals' instead of `portfolios'), are exploited for classifying the sentence's risk impact (similar to the concepts used in self-interpretable image classification -- e.g.,~\cite{turbe2024protos}). 
\textit{Explainability} explicates the relevance of `rearing animals' to sustainability risk, further justifying the model's classification. It explains the link between `rearing animals', deforestation, methane emissions, and strained water resources, resolving the ambiguity surrounding the term `rearing animals'. \textit{Faithfulness} ensures that the insights from \textit{interpretability} and \textit{explainability} are meaningful in that they accurately reflect the model's mechanism. Put together, \textit{interpretability, explainability} and \textit{faithfulness} prove the robustness of the model's decision-making process, increasing an analyst's trust in its classification output. 


Integrating these capabilities within NLP has already proven to be of significant for trust-dependent sectors like healthcare~\cite{han2022hierarchical}. This reinforces their broader applicability to corporate sustainability analysis, where trustworthy and credible NLP insights are also critical~\cite{guix2022trustworthy}. Such integration will allow NLP models to produce more actionable insights for analysts. 


\subsection{Inferring patterns for greenwashing research}

\textit{Interpretable} NLP models can afford cues on the interdependence between different features within sustainability disclosures. 
This can inspire researchers to theorise and subsequently validate the causal and correlation relationships between attributes relevant to corporate sustainability analysis~\cite{lipton2018mythos,mengaldo2024explain}. 
By doing so, our understanding of under-explored issues can be augmented, mitigating longstanding issues within the domain.

To elaborate, white-box models or intrinsically \textit{interpretable} models (Naive Bayes Classifiers, Generalised Additive Models, Decision Trees etc.), allow us to grasp the decision-making process from feature inputs to classification, implying the significance of specific features and their relationships with predictions. For instance, text classification models designed with \textit{interpretability} techniques can elucidate the influence of keywords for classification decisions~\cite{zhang2023neurosymbolic}. Developing a similar white-box model for classifying greenwashing texts allows us to study how specific text features result in a positive classification. This can offer hints on the semantic and syntactic features that characterise greenwashed texts. From a `machine' based perspective, it can powerfully reveal latent data patterns easy for an analyst to miss. This fresh angle would strengthen existing greenwashing research that predominantly involves human analysis of sustainability texts~\cite{de2020concepts}, making greenwashing more understandable and detectable. In this fashion, \textit{interpretability} can improve the \textit{unreliability} and \textit{ambiguity} issues in the field arising from the greenwashing phenomenon. 


\subsection{Generalisability across frameworks and sources}
 Sustainability disclosures stem from different sources and are presented through diverse frameworks. As such, they can have differing linguistic and textual features, posing a challenge for NLP models to generalise. To elaborate, while firms can focus corporate sustainability disclosures on positive efforts~\cite{Cho2010}, government-linked sustainability disclosures may mandate firms to disclose sustainability related risks (such as U.S. Securities and Exchange Commission 10k reports)~\cite{kim2023real}. An \textit{impact classification} model trained on corporate sustainability disclosures may not adapt well to government-linked sustainability disclosures, as the latter more frequently carries negative impact. On top of heterogeneous sustainability data, the \textit{generalisability} problem is further compounded by the lack of available datasets within the field~\cite{LEE2023119726}. This reduces the diversity of data for training robust models. In light of limited data availability, \textit{interpretability} methods provides an alternative means for mitigating generalisability issues. 


Specifically, \textit{interpretability} allows us to understand a model's sensitivity to training data, providing insight into portions of data that cause overfitting. Recent works such as~\cite{nickl2023memory} detail a method to do so through a memory perturbation equation. The paper faithfully derives `shirt, pullover' as classes the model is most sensitive to while training on the FMNIST dataset, and demonstrates how removing these classes can improve model generalisability performance. Moreover, it also highlights how specific samples and training epochs impact model sensitivity. While this method has been deployed for computer vision, the same principles can be adapted to NLP for sustainability analysis. Leveraging such a tool, NLP practitioners can consider removing portions of data most prone to model overfitting. This optimises training for general performance across the different sustainability disclosures. By enhancing generalisability, NLP models can effectively analyse diverse types of information associated with a firm's sustainability efforts~\cite{LEE2023119726}. This potentially compensates for \textit{incompleteness} and \textit{unreliability} data problems, by reducing over-reliance on specific sources that omit important details or incredible information. 


\subsection{Mitigating bias}
By producing consistent and reproducible insights, NLP can partially address the \textit{potential bias} of analysts that conduct corporate sustainability analysis. However, NLP also risks introducing its own biases if not designed transparently. To elaborate, NLP constructed with XAI capabilities can elucidate its decision-making process, allowing analysts to verify its learned features~\cite{raghavan2020mitigating}. While this may still incur the partiality of analyst judgement, an XAI enhanced model can conversely highlight potentially overlooked features to analysts, guiding them to produce more balanced corporate sustainability assessments. Consequently, XAI capabilities can enable a synergistic interaction between NLP models and analysts for reducing potential bias.

\section{Conclusion}
XNLP algorithms are well equipped for sustainability analysis given that they encompass both linguistic understanding and XAI capabilities. To elaborate on this, while lexical, semantic and syntactic analysis of text automates and enhances sustainability analysis, XAI further mitigates the subjectivity challenge of sustainability assessments. However, NLP for sustainability analysis has not yet embraced these directions, leaving its vast potential untapped. This underscores an opportunity for researchers and developers to explore the strategies highlighted---achieving trustworthy insights, inferring patterns for greenwashing research, generalisability across frameworks and sources, mitigating bias---to enhance NLP for tasks (T1-T5). These efforts can not only improve sustainability analysis but potentially reshape how businesses, investors, and regulators address corporate sustainability reporting and attribution. \\

\noindent\textbf{Limitations} \\
Given that the development of NLP methods for sustainability analysis is still nascent, several papers that describe NLP methods for sustainability analysis entail emerging research that have not yet undergone extensive peer-review. Despite our attempts to filter out papers with evident shortcomings, it is important to acknowledge that some of the discussed papers may contain non-authoritative viewpoints. \\


\bibliographystyle{plain}
\bibliography{main}

\end{document}